\documentclass{svproc}
\usepackage{url}
\usepackage{graphicx}
\usepackage{verbatim}
\usepackage{tabularx}
\usepackage{hyperref}
\usepackage{booktabs}
\usepackage{cite}
\usepackage{float} 
\usepackage{amsmath}

\begin{document}
\mainmatter              
\title{A Framework for Assessing Sustainability Conflicts in the Design of Medical Devices}
\titlerunning{Framework for Sustainability Conflicts}  
%
\author{Apala Chakrabarti}
%
\authorrunning{Apala Chakrabarti} 
%
%
\institute{Centre of Excellence in Design, Indian Institute of Science, Bengaluru, India \\
\email{apala@fsid-iisc.in}}

\maketitle              

\begin{abstract}
Medical devices improve healthcare outcomes but often involve sustainability conflicts across environmental, economic, and social pillars. Existing approaches typically prioritize one or two pillars, lacking a unified framework to assess cross-domain conflicts. This paper presents a structured framework to identify and quantify sustainability conflicts in medical device design. It integrates life cycle analysis, cause-effect (CE) mapping, and Multi-Criteria Decision Analysis (MCDA) to evaluate the impact of design choices across all three pillars. A case study of an oxygen concentrator illustrates the framework’s application, generating a composite sustainability score based on identified trade-offs. The framework supports informed, data-driven design decisions while meeting regulatory and ethical requirements. This work addresses a key gap in sustainable medical device development by offering a repeatable, quantifiable approach to conflict assessment and resolution.

\end{abstract}
\section{Introduction}

Sustainability in medical device design is a complex yet urgent challenge. The healthcare sector contributes approximately 5\% of global greenhouse gas emissions \cite{msu2022}, and medical devices generate significant waste—particularly from single-use components—while the global market continues to expand, projected to reach \$678.9 billion by 2025 \cite{arterex2025}. This rapid growth underscores the need to address environmental burdens without compromising safety, affordability, and accessibility.

Efforts to reduce environmental impacts—such as eco-design strategies or the adoption of biodegradable materials—often conflict with economic viability and stringent safety requirements. A conflict, in this context, occurs when fulfilling one sustainability objective undermines another \cite{conflict}. Social considerations, such as patient safety, usability, and ethical obligations, introduce further trade-offs. Yet most existing sustainability assessments evaluate each dimension in isolation, failing to capture these interdependencies. This motivates the need for an integrated approach to assess and balance sustainability challenges in medical device design.

\section{Background Literature}

\par A number of standards and regulations already promote sustainability in healthcare technologies. For example, IEC 60601-1-9 and directives such as WEEE and REACH address energy efficiency, waste reduction, and public health \cite{weeeDirective, reachRegulation}, but they focus on specific aspects and lack integrated methods for assessing trade-offs across multiple sustainability pillars.  

\par ISO 14971 and ISO 13485 govern risk management and compliance for medical devices \cite{iso14971, iso13485}, yet omit mechanisms for evaluating conflicts between environmental, economic, and social objectives. Broader tools such as life cycle assessment (LCA) provide structured environmental evaluation but overlook social and economic dimensions \cite{hauschild2017}. Prior studies also note that the absence of unified frameworks constrains holistic sustainability assessment in design \cite{moultrie2015}.

\section{Challenges and Constraints in Medical Device Design}

Medical devices face design challenges distinct from other industries, arising from their critical role in healthcare and strict regulatory oversight. While general sustainability frameworks emphasise environmental and cost optimisation, medical devices must also prioritise patient safety, reliability, and ethical obligations \cite{iso14971}. These priorities create constraints that complicate the adoption of circular or resource-efficient design practices.

\begin{table}[H]
\caption{Sustainability-Relevant Aspects in Medical Devices vs. Other Industries}
\label{tab:medvot}
\begin{center}
\resizebox{\textwidth}{!}{
\begin{tabular}{r@{\hspace{0.6em}}p{5cm}@{\hspace{1.2em}}p{5cm}}
\hline
\multicolumn{1}{l}{\rule{0pt}{12pt} \textbf{Aspect}} & \textbf{Medical Devices} & \textbf{Other Industries (excluding nuclear and aerospace sectors)} \\[2pt]
\hline\rule{0pt}{12pt}
Safety/Efficacy        & Paramount, legally mandated                      & Important, but less regulated \\
Regulatory Barriers    & Highly regulated, many exemptions from general sustainability laws & Fewer exemptions, broader applicability \\
Waste Management       & Biohazardous, specialized disposal              & Standard recycling/refurbishment \\
Lifecycle Traceability & Mandatory, includes disposal instructions       & Rarely required \\
Innovation Speed       & Slow, due to validation and liability           & Generally faster, fewer constraints \\
Supply Chain           & Highly controlled, clinical-grade materials     & More flexibility \\
Circular Economy       & Limited by contamination and safety             & Potential to be widely implemented \\[2pt]
\hline
\end{tabular}
}
\end{center}
\end{table}

Ethical obligations such as informed consent, post-market surveillance, and usability for diverse patient groups enhance social sustainability but often increase costs and resource requirements. Sterility and single-use mandates reduce opportunities for reuse or remanufacturing, while long regulatory timelines delay market entry \cite{iso13485}. In addition, reliance on clinical-grade materials and extensive risk-mitigation protocols imposes further restrictions on innovation speed and circular economy adoption. These challenges are summarised in Table~\ref{tab:medvot}, which compares sustainability-relevant aspects in medical devices with those in other industries. Attributes unique to medical devices are further detailed in Table~\ref{tab:pillarattributes}, forming the basis for conflict identification in this study.
 \setlength{\intextsep}{6pt}
\begin{table}[H]
\caption{Pillar-Specific Attributes Unique to Medical Devices}
\label{tab:pillarattributes}
\begin{center}
\resizebox{\textwidth}{!}{
\begin{tabular}{p{5cm}@{\hspace{1.2em}}p{5cm}@{\hspace{1.2em}}p{5cm}}
\hline
\textbf{Environmental} & \textbf{Economic} & \textbf{Social} \\[2pt]
\hline\rule{0pt}{12pt}
Biohazardous waste requiring specialized disposal & High regulatory compliance costs across lifecycle & Patient safety validated through clinical evidence \\
Sterility requirements limit reuse/remanufacturing & Prolonged time-to-market due to validation & Usability for diverse users in critical settings \\
Material selection constrained by biocompatibility & Market access linked to reimbursement frameworks & Informed consent and ethical oversight mandatory \\
Single-use design driven by infection control & High cost of liability and risk mitigation & Designs must minimize risk of user error \\
\hline
\end{tabular}
}
\end{center}
\end{table}
\vspace*{-10pt} 
 \setlength{\textfloatsep}{6pt}
\section{Methodology}

\par This study adopts the Design Research Methodology (DRM) \cite{chakrabartiDRM} as its guiding approach, following a Type-4 structure. In the Research Clarification (RC) stage, gaps in existing sustainability frameworks and conflict assessment methods were identified. The Descriptive Study 1 (DS-1) stage involved a comprehensive literature review covering sustainability standards, medical device design requirements, and domain-specific constraints. In the Prescriptive Study (PS) stage, a structured framework was developed for identifying, categorizing, and scoring sustainability conflicts using MCDA. Finally, Descriptive Study 2 (DS-2) tested the framework on a case study of an oxygen concentrator to evaluate its applicability.

\begin{figure}[H]
    \centering
    \includegraphics[width=0.7\linewidth]{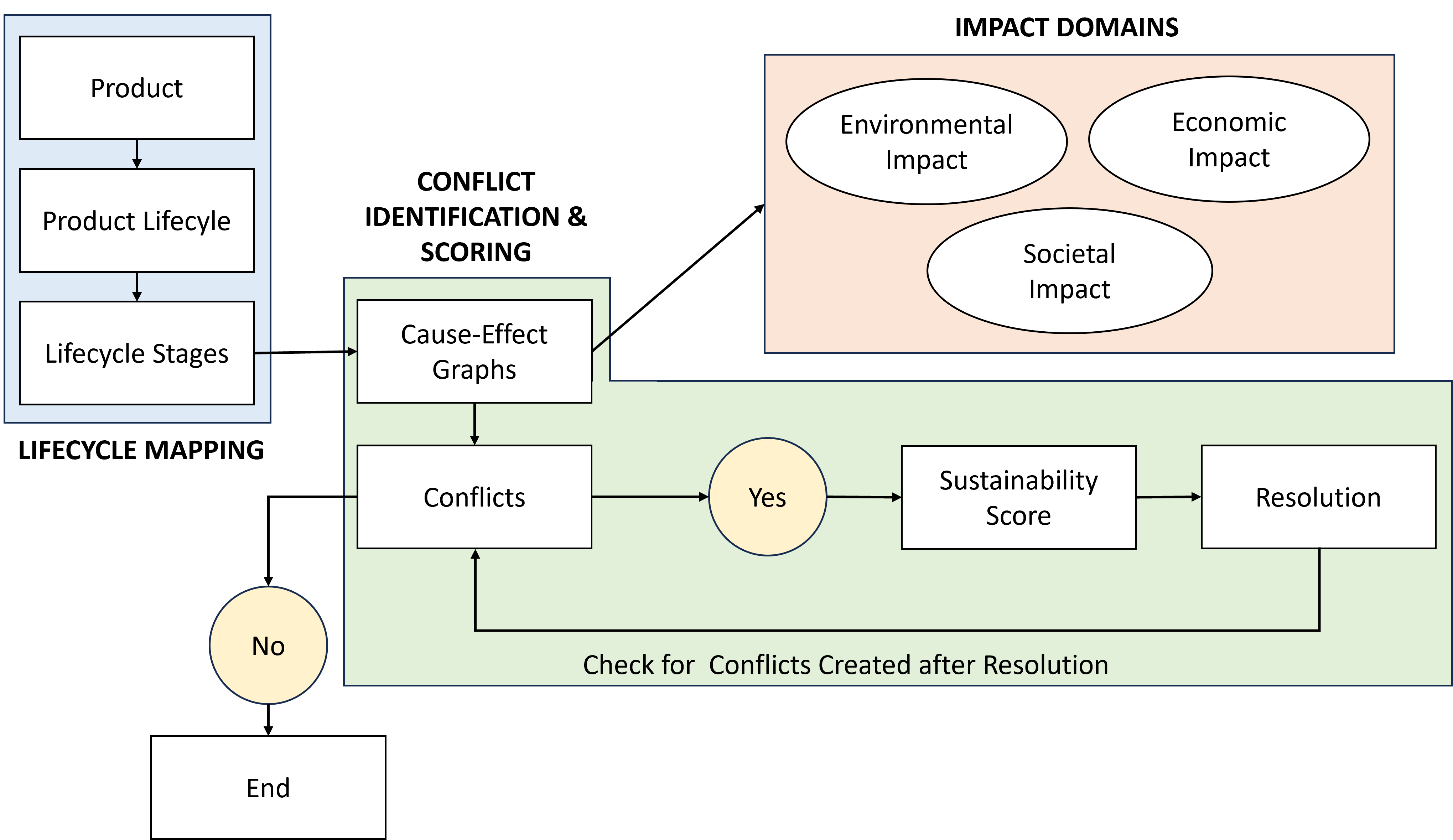}
    \caption{Process flowchart for sustainability assessment.}
    \label{fig:process_flow}
\end{figure}

\par The proposed framework comprises three main components: (1) product life cycle and cause-effect analysis, (2) conflict database construction, and (3) multi-criteria decision analysis (MCDA) for sustainability scoring. The overall iterative process is illustrated in Figure~\ref{fig:process_flow}.

\subsection{Data Structuring and Product Life Cycle Mapping}

\par The framework begins with a structured analysis of the product life cycle (PLC). Each product is decomposed into its 5 key life cycle stages: raw material acquisition, manufacturing, transportation, usage, and end-of-life management.

\par The PLC stages used in this study were derived from publicly available sources, including product documentation, manufacturer manuals, technical blogs, and product specifications. These were cross-checked against relevant standards (ISO 14971, ISO 13485, WEEE, and REACH) and prior studies to ensure consistency. 

\par Cause-effect (CE) mapping is conducted at each life cycle stage to identify relationships between design or process decisions and sustainability impacts. In the cause-effect graph, a conflict is identified when a single cause node (typically representing a design decision or intervention) leads to multiple downstream effects, each assessed independently across sustainability domains—environmental, economic, and social.

\par Each effect is evaluated as having either a positive or negative impact with respect to its associated sustainability objective. If all downstream effects of a cause are uniformly positive or uniformly negative, no conflict is recorded. 

\par For example, consider the adoption of a biodegradable polymer in device packaging. This design choice (cause) may lead to:
(a) a reduction in long-term environmental impact (positive environmental effect),
(b) a higher unit cost of production (negative economic effect), and
(c) increased consumer acceptance due to eco-labeling (positive social effect). Because the effects span both positive and negative outcomes across domains, the design choice presents a conflict. 

\subsection{Conflict Database Construction}

\par Conflicts identified through cause-effect mapping are organized into a structured database. Each entry captures the relevant life cycle stage, involved sustainability domains, design choice, and its associated trade-offs. 
\vspace{\baselineskip}

\par The database is populated after creating CE graphs for various products and identifying conflicts. Conflicts are recorded only where a clear conflict between at least two sustainability domains is identified.

\subsection{Multi-Criteria Decision Analysis (MCDA) Scoring Framework}

This section introduces a multi-criteria scoring method that incorporates both the impact of sustainability effects and the relative importance (called 'weight', see below) of each pillar—social, environmental, and economic—within medical device design.

\subsection{Classification of Strength of Impact}

Each effect in the cause-effect graph is evaluated for its impact using a three-tiered scale:

\begin{itemize}
    \item \textbf{High (0.75)}: Irreversible or cross-system impacts requiring urgent mitigation.
    \begin{itemize}
        \item Examples: Biohazard disposal failures, non-recyclable rare earth mining, labor exploitation, high energy sterilization, or regulatory violations.
    \end{itemize}

    \item \textbf{Medium (0.50)}: Containable effects that can escalate cumulatively.
    \begin{itemize}
        \item Examples: Excess packaging, elevated manufacturing costs, above-benchmark emissions, or usability-driven inefficiencies.
    \end{itemize}

    \item \textbf{Low (0.25)}: Localized, reversible impacts with minimal systemic risk.
    \begin{itemize}
        \item Examples: Slight material overuse, basic plastics, minor delays, or operational noise.
    \end{itemize}
\end{itemize}

\subsection{Scoring Method}

Each effect is scored as:

\[
\text{Effect Score} = \text{impact} \times \text{Pillar Weight}
\]

Positive and negative effects are scored separately. Aggregation yields:

\begin{itemize}
    \item \( P \): Total positive score
    \item \( N \): Total negative score
    \item \( T = P + N \): Total weighted impact
    \item \( R = N / T \): Sustainability score
\end{itemize}

\subsection{Sustainability Score Categories}

\par The sustainability score is interpreted using Table~\ref{tab:score_categories}, which classifies the score into five categories. These categories reflect the overall impact of negative sustainability impacts accumulated across all stages of the product life cycle.

\begin{table}[H]
\caption{Sustainability classification based onthe sustainability score}
\label{tab:score_categories}
\begin{center}
\begin{tabular}{p{3cm} p{4cm} p{6cm}}
\hline
\textbf{Sustainability Score}& \textbf{Category} & \textbf{Description} \\
\hline
0.00 & Fully Sustainable & All effects are beneficial; no sustainability conflicts. \\
0.01 -- 0.33 & Highly Sustainable & Minor trade-offs; benefits outweigh drawbacks. \\
0.34 -- 0.66 & Moderately Sustainable & Balanced impacts; moderate design concerns. \\
0.67 -- 0.99 & Highly Unsustainable & Negative effects dominate; mitigation needed. \\
1.00 & Fully Unsustainable & Entirely negative; unsuitable configuration. \\
\hline
\end{tabular}
\end{center}
\end{table}
\vspace{-8pt}
\setlength{\textfloatsep}{6pt}
\setlength{\intextsep}{6pt} 
\subsection{Conflict Magnitude}

The total impact score \( T = P + N \) represents the cumulative weighted strength of all sustainability-relevant effects—both positive and negative—associated with a product. This score serves two purposes:

\begin{enumerate}
    \item \textbf{Conflict Count}: By evaluating the number of distinct conflicts, the framework captures the breadth of trade-offs present in the product design.
    
    \item \textbf{Conflict Intensity}: The value of \( T \) quantifies the aggregate weighted impact of all identified conflicts. It is derived from the combined impact and priority of both positive and negative sustainability effects. A higher \( T \) value indicates that a product involves more numerous or more severe sustainability trade-offs. This makes \( T \) a proxy for the total sustainability burden of a design. By comparing the \( T \) scores of different products, stakeholders can assess which product exhibits higher sustainability related interactions, even if their sustainability ratios \( R \) (i.e., balance of negative to total impact) are similar.
\end{enumerate}

\par Together, these metrics support both intra-product analysis (impact within a product) and inter-product benchmarking (comparison between devices).

\section{Case Study: Sustainability Assessment of an Oxygen Concentrator}

To illustrate the proposed framework, a case study is conducted on a medical-grade oxygen concentrator. The device extracts and delivers concentrated oxygen from ambient air and is commonly used in hospitals and home-care environments.

\subsection{Product Life Cycle and Cause–Effect Mapping}

\subsubsection{Life Cycle Stage Mapping}

The oxygen concentrator was analyzed across five product life cycle stages:

\begin{enumerate}
    \item \textbf{Raw Material Acquisition}: Extraction and processing of zeolite (for oxygen separation), polymers, and electronic materials.
    \item \textbf{Manufacturing}: Component fabrication, device assembly, and packaging.
    \item \textbf{Transportation}: Long-distance distribution to hospitals and home-care facilities.
    \item \textbf{Use Phase}: Continuous electrical operation for oxygen delivery to patients.
    \item \textbf{End-of-Life}: Disposal of spent filters, batteries, plastic housing, and circuit boards.
\end{enumerate}

\subsubsection{Cause–Effect Graph}

\begin{figure}[H]
    \centering
    \includegraphics[width=\linewidth]{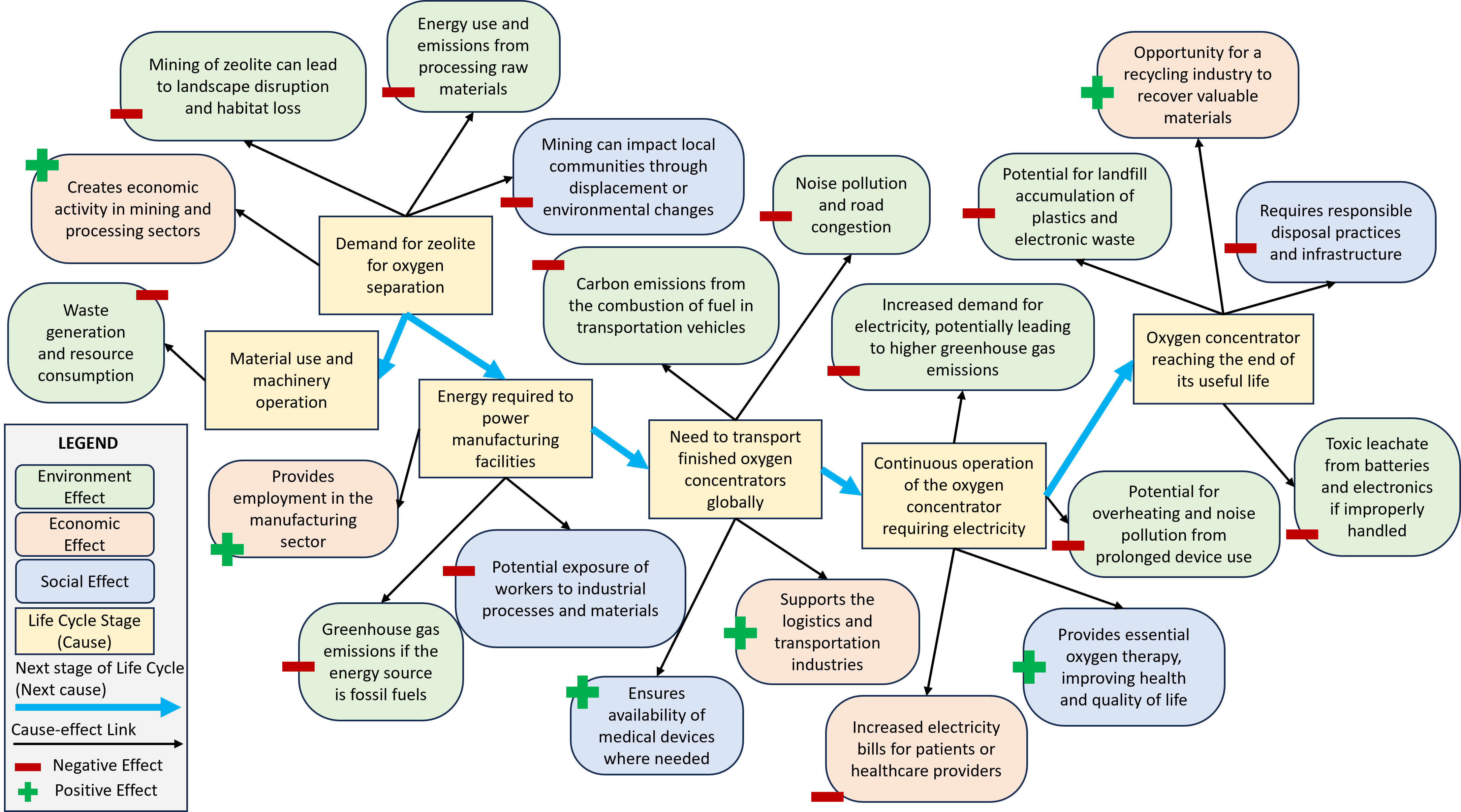}
    \caption{Cause-effect graph of oxygen concentrator across life cycle stages.}
    \label{fig:ce_oxy}
\end{figure}

A cause-effect (CE) graph was developed to represent sustainability-relevant effects across each life cycle stage (Figure~\ref{fig:ce_oxy}). Nodes represent design choices and downstream effects; edges denote causal links. Visual arrow orientation is solely for layout clarity.

\subsection{Conflict Identification and Database Construction}

Sustainability conflicts were derived from the CE graph by identifying trade-offs between the three pillars. Table~\ref{tab:conflict_entries} summarizes the five key conflicts selected.

\begin{table}[H]
\vspace{-0.8em}
\caption{Summary of sustainability conflicts in the oxygen concentrator}
\label{tab:conflict_entries}
\begin{center}
\resizebox{\linewidth}{!}{
\begin{tabular}{l p{2.9cm} p{2.9cm} p{2.9cm} p{2.9cm} p{2.9cm}}
\hline
\textbf{Attribute} & \textbf{Conflict 1: Raw Material} & \textbf{Conflict 2: Use Phase} & \textbf{Conflict 3: End-of-Life} & \textbf{Conflict 4: Manufacturing} & \textbf{Conflict 5: Transport} \\
\hline
Conflict ID & C-01 & C-02 & C-03 & C-04 & C-05 \\
Sustainability Domains & Environmental vs Economic, Social & Environmental vs Economic, Social & Social vs Environmental, Economic & Environmental vs Economic & Environmental vs Social \\
Life Cycle Stage & Raw Material Acquisition & Use Phase & End-of-Life & Manufacturing & Transportation \\
Design Choice & Zeolite for O\textsubscript{2} separation & Continuous electric operation & Single-use filters and electronics & Energy-intensive assembly & Long-distance shipment \\
Positive Impact & Low cost; high O\textsubscript{2} efficiency & Reliable oxygen delivery & Infection control; clinical hygiene & Job creation; device reliability & Global accessibility; logistic coverage \\
Negative Impact & Emissions; habitat loss; material inaccessibility & High power use; overheating risk & E-waste; toxic leachate; disposal cost & Emissions; material waste & Carbon emissions; noise; congestion \\
\hline
\end{tabular}
}
\end{center}
\end{table}
\setlength{\textfloatsep}{6pt}
\setlength{\intextsep}{6pt} 

\subsection{Sustainability Scoring via MCDA}

\subsubsection{Pillar Weights:}

Weights for the environmental, economic, and social pillars were assigned using a qualitative pairwise prioritization based on sustainability objectives relevant to medical device design. However, it is acknowledged that affordability directly affects accessibility and, by extension, patient safety. This interdependence is addressed through conflict scoring, even if its weight is relatively lower in the final aggregation.
 \setlength{\intextsep}{5pt}
\begin{itemize}
    \item \textbf{Social (0.75)}: Patient safety, equity, and usability.
    \item \textbf{Environmental (0.50)}: Resource use, emissions, waste.
    \item \textbf{Economic (0.25)}: Cost, supply, efficiency.
\end{itemize}
 \setlength{\intextsep}{5pt}
\subsubsection{Conflict Impact Scoring:}

Each conflict was scored based on CE graph analysis using a three-tiered scale derived from impact strength as seen in Table \ref{tab:negative_conflict_scores}:
 \setlength{\intextsep}{5pt}
\begin{itemize}
    \item \textbf{High (0.75)}: Irreversible or cross-system impacts requiring urgent mitigation.
    \item \textbf{Medium (0.50)}: Containable effects that can escalate cumulatively.
    \item \textbf{Low (0.25)}: Localized, reversible impacts with minimal systemic risk.
\end{itemize}
 \setlength{\intextsep}{5pt}
Table \ref{tab:negative_conflict_scores} presents the negative impact scores across sustainability pillars for each conflict. For conflicts such as C-01 and C-05, the environmental impact score of 1.50 is the sum of two distinct environmental effects, each contributing 0.75 to the total. This reflects multiple negative environmental consequences associated with a single design choice.

\begin{table}[H]
\caption{Negative impact scores across sustainability pillars}
\label{tab:negative_conflict_scores}
\begin{center}
\begin{tabular}{l c c c}
\hline
\textbf{Conflict ID} & \textbf{Environmental} & \textbf{Economic} & \textbf{Social} \\
\hline
C-01 & 1.50 & 0 & 0.75 \\
C-02 & 0.50 & 0 & 0.50 \\
C-03 & 1.00 & 0 & 0 \\
C-04 & 0.75 & 0.50 & 0 \\
C-05 & 1.50 & 0 & 0.50 \\
\hline
\textbf{Total} & \textbf{5.25} & \textbf{0.50} & \textbf{1.75} \\
\hline
\end{tabular}
\end{center}
\end{table}
\vspace{-0.8em}
Total negative score \( N \) is given by:

\begin{equation}
N = (5.25 \times 0.5) + (0.50 \times 0.25) + (1.75 \times 0.75) =\boxed{4.0625}
\end{equation}

Similarly, the total positive score is computed (Table~\ref{tab:positive_conflict_scores}).

\begin{table}[H]
\caption{Positive impact scores across sustainability pillars}
\label{tab:positive_conflict_scores}
\begin{center}
\begin{tabular}{l c c c}
\hline
\textbf{Conflict ID} & \textbf{Environmental} & \textbf{Economic} & \textbf{Social} \\
\hline
C-01 & 0 & 0.75 & 0 \\
C-02 & 0 & 0.75 & 0 \\
C-03 & 0 & 0.50 & 0.75 \\
C-04 & 0 & 0 & 0.75 \\
C-05 & 0 & 0.50 & 0 \\
\hline
\textbf{Total} & \textbf{0} & \textbf{2.50} & \textbf{1.50} \\
\hline
\end{tabular}
\end{center}
\end{table}

The total positive score \( P \) is given by:

\begin{equation}
P = (0 \times 0.5) + (2.5 \times 0.25) + (1.5 \times 0.75) = \boxed{1.75}
\end{equation}

Total impact score \( T \) is computed as:

\begin{equation}
T = 1.75 + 4.0625 = \boxed{5.8125}
\end{equation}

This value represents the aggregate weighted effect of all sustainability-relevant outcomes across the product life cycle. The sustainability score \( R \) is computed as:

\begin{equation}
R = \frac{4.0625}{5.8125} = \boxed{0.6982}
\end{equation}

\subsection{Inference Based on Scoring Results}

\subsubsection{Conflict Intensity}

The total impact score, \( T = 5.8125 \), quantifies the combined impact and weighted influence of all identified sustainability conflicts. This value reflects the overall sustainability burden of the product design, capturing both beneficial and adverse effects across the product life cycle.

\subsubsection{Sustainability Distribution}

The sustainability ratio \( R = 0.6982 \) falls outside the range of 0.34–0.66, and thus corresponds to the category: Highly Unsustainable

\par This classification indicates that negative effects, concentrated in the environmental and social pillars, dominate the product’s profile. The unsustainability arises mainly from high electricity demand, e-waste from single-use filters, and material-related emissions. Potential strategies include improving energy efficiency, redesigning filters for reuse or recyclability, and using lower-impact materials.

\section{Discussions, Conclusions and Future Work}

\par This paper proposed a structured framework to identify and quantify sustainability conflicts in medical device design. By combining life cycle mapping, cause–effect analysis, and MCDA, it enables evaluation across environmental, economic, and social pillars. The method distinguishes positive and negative effects, aggregates them into a composite score, and supports early design decisions. Pillar weights are configurable to reflect contextual priorities, and regulatory and patient-related requirements are embedded through existing standards. The framework also accommodates provisional early-stage data, updated iteratively as designs mature.  

\par Limitations include manual CE mapping, reliance on expert scoring, and simplified treatment of interdependencies. Future work will pursue AI-assisted automation, including large language models for scalability and data generation under uncertainty. Broader validation, refinement of scoring criteria, and integration with design tools will be undertaken, while extended CE graphs will address non-linear interactions. These improvements aim to increase robustness, comparability, and practical applicability of the framework. Future refinements will include clearer, standardised criteria for impact assignment and weight selection to reduce subjectivity. In addition, while regulatory requirements are embedded through existing standards, incorporating direct input from patients and regulators may further strengthen the social assessment dimension.

%
%

\end{document}